\documentclass[12pt]{iopart}
\usepackage{amssymb,setstack}

\def\ba{\begin{array}}
\def\ea{\end{array}}
\def\be{\begin{equation}}
\def\ee{\end{equation}}

\def\sp{{\rm span}}

\def\C{{\mathbb C}}

\def\N{{\mathbb N}}

\def\g{{\mathfrak g}}
\def\h{{\mathfrak h}}
\def\l{{\mathfrak l}}
\def\m{{\mathfrak m}}
\def\n{{\mathfrak n}}
\def\r{{\mathfrak r}}
\def\p{{\mathfrak p}}
\def\q{{\mathfrak q}}
\def\s{{\mathfrak s}}
\def\sl{{\mathfrak s \mathfrak l}}
\def\so{{\mathfrak s \mathfrak o}}

\def\z{{\mathfrak z}}
\def\n{{\mathfrak n}}
\def\der{{\mathfrak{Der}}}
\def\inn{{\mathfrak{Inn}}}

\def\span{{\rm span}}

\def\NR{{\rm NR}}
\def\ad{{\rm ad}}

\def\mod{{\rm \, mod \,}}

\def\cent{{\mathfrak c \mathfrak e \mathfrak n \mathfrak t}}

\begin{document}

\newtheorem{theorem}{Theorem}
\newtheorem{prop}{Proposition}
\newtheorem{conjecture}{Conjecture}

\title[Structure of solvable and Levi extensions of nilpotent algebras]{On the structure of maximal solvable extensions and of Levi extensions of nilpotent algebras}

\author{L \v Snobl\dag}

\address{\dag\ Department of Physics,  
Faculty of Nuclear Sciences and Physical Engineering, 
Czech Technical University in Prague, B\v rehov\'a 7, 115 19 Prague 1, Czech Republic }

\eads{\mailto{Libor.Snobl@fjfi.cvut.cz}}

\begin{abstract}
We establish an improved upper estimate on dimension of any solvable algebra $\s$ with its nilradical isomorphic to a  given nilpotent Lie algebra $\n$. Next we consider Levi decomposable algebras with a given nilradical $\n$ and investigate restrictions on possible Levi factors originating from the structure of characteristic ideals of $\n$. We present a new perspective on Turkowski's classification of Levi decomposable algebras up to dimension 9.
\end{abstract}

\pacs{02.20.Sv,02.20.Qs}

\ams{17B30,17B05,17B81}

\section{Introduction}\label{intro}

The aim of the present paper is to establish some general properties of solvable and Levi extensions of nilpotent Lie algebras.

As is well known, the problem of classification of all solvable (including nilpotent) Lie algebras in an arbitrarily large finite dimension is presently unsolved and is generally believed to be unsolvable; at least unless some completely new ideas emerge and a new understanding of the notion ``classification'' itself develops. The problem stems from an obvious fact that the number of solvable Lie algebras in higher dimensions increases drastically, and infinite parametrized families of such nonisomorphic algebras arise already in very low dimensions. This behavior is in a stark contrast with the theory of semisimple algebras where only finitely many algebras exist in any given dimension and their full classification was completed already long time ago \cite{Cartan,Gan}. Because any Lie algebra is a semidirect sum of its maximal solvable ideal and a semisimple subalgebra by theorem of Levi \cite{Levi}, the difficulty in the classification of solvable algebras also shows up in the classification of all types of non--semisimple algebras.

All known full classifications of non--semisimple algebras terminate at relatively low dimensions. First low dimensional classifications were established already by Lie and his contemporaries in \cite{Lie,LieScheffers,Umlauf}, their results are reviewed e.g. in \cite{PSW}. Newer results since mid--20th century are the classifications of nilpotent algebras in dimension 6 \cite{Morozov}, dimension 7 \cite{Saf,Seeley,GKh1,Gong}, in dimension 8 \cite{Tsa1} and partially in dimension 9 \cite{Tsa2}, and of solvable algebras in dimension 5 \cite{Mub2}, dimension 6 \cite{Mub3,Tur2} and some partial results in dimension 7 \cite{Parry}. Algebras of semidirect sum type, i.e. Levi decomposable algebras, were classified up to dimension 8 in \cite{Tur1} and in dimension 9 in \cite{Tur4}.

As a possible stopgap solution, the idea of a classification of solvable extensions of certain particular classes of nilpotent Lie algebras, i.e. of all solvable, non--nilpotent algebras with the given nilradical, of arbitrarily large dimension emerged. It is based on a belief that the knowledge of full classification of all solvable extensions of certain series of nilradicals can be very useful for both theoretical considerations -- e.g. testing various hypotheses concerning the general structure of solvable Lie algebras -- and practical purposes -- e.g. when a generalization of a given algebra or its nilradical to higher dimensions is needed in some physical situation. Such need arises for example in the construction of superintegrable systems from a given solvable Lie algebra and its Casimir invariants which was introduced in \cite{BB}. Another application comes from the construction of cosmological models in higher dimensions, now fashionable e.g. in string cosmology, using algebraic methods \cite{Petrov,KSHMC}. Lie algebras of Killing vectors are in many cases solvable\footnote{It was probably the first appearance of solvable groups and algebras in physics other than their use in solution of ODEs using the reduction method of S. Lie.} as was realized already in \cite{Landau} using the classification of homogeneous spaces  \cite{Bianchi}. Higher dimensional solvable Lie algebras and their semidirect sums with semisimple algebras therefore appear naturally in such constructions in higher spacetime dimensions, see e.g. \cite{Tur3}. If some of the properties of the resulting spacetimes ought to resemble the behavior of their low--dimensional counterparts then it is natural to expect that also their algebras of Killing vectors should have some properties in common, e.g. to belong to one common series of algebras.

Gradually a series of classifications of solvable extensions was performed. The first one was done by P. Winternitz together with J. L. Rubin in \cite{RW}. The series then continued throughout the years in \cite{NW,TW,SW,SW1,SK}. All these papers dealt with the problem of classification of all solvable Lie algebras with the given $n$--dimensional nilradical, e.g. Abelian algebra, Heisenberg algebra, the algebra of strictly upper triangular matrices etc., for arbitrary finite dimension $n$. Similar sequences have been recently investigated also by other research groups in \cite{ACSV} (naturally graded nilradicals with maximal nilindex and a Heisenberg subalgebra of codimension one) and \cite{WLD} (a certain sequence of quasi--filiform  nilradicals). A recent paper \cite{CS} generalized results of \cite{SW,SW1} in the sense that all solvable extensions of $\N$--graded filiform nilradicals were classified.

Levi decomposable algebras with a fixed structure of their nilradical were considered in \cite{RCSheis} for Heisenberg nilradicals. 
\medskip

The present article builds on our experience gained in \cite{SW,SW1,SK}. In those papers we analyzed in detail the structure of solvable extensions of particular chosen sequences of nilradicals. Here, we use methods and ideas developed in \cite{SW,SW1} in a different direction, namely, to give at least some general estimate on dimension of any solvable Lie algebra with a given nilradical.

Next we investigate Levi decomposable algebras with a given nilradical. Our main concern is to formulate some necessary conditions for the existence of nontrivial Levi extension(s).  We formulate these conditions in terms of  characteristic ideals of the nilradical, in particular in terms of lower central series. 
\medskip

The structure of our paper is as follows. After an introduction of notation and a brief review of known facts in Sections \ref{prelim} and \ref{introsan} we present an improved upper bound on dimension of any solvable Lie algebra with the given nilradical in Section \ref{upperbound}. In Section \ref{levinilr} we study the structure of Levi decomposable algebras with the given nilradicals and provide a novel perspective on the classification of P. Turkowski. Finally, we introduce some open questions in Conclusions.

Throughout the paper the analysis is done over the fields of complex and real numbers unless indicated otherwise.

\section{Notation}\label{prelim}
We shall often need to refer to the Jacobi identity 
\begin{equation}\label{JI}
 [x,[y,z]]+ [y,[z,x]] + [z,[x,y]]=0
\end{equation} 
for a particular triple $x,y,z$ of vectors in $\g$. For brevity, we speak about the Jacobi identity $(x,y,z)$ in such case. A Lie bracket of two vector subspaces is defined to be the whole span
$$ [\h,\tilde \h]=\span \{ [x,\tilde x] | x \in \h, \tilde x \in \tilde \h \}.$$

For a given Lie algebra $\g$ we consider the following three series of ideals.

The \emph{derived series} 
$ \g = \g^{(0)} \supset \g^{(1)} \supset \ldots \supset \g^{(k)} \supset \ldots $ 
is defined recursively
\begin{equation}
 \g^{(0)}=\g, \qquad  \g^{(k)} = [\g^{(k-1)},\g^{(k-1)}], \ k\geq 1.
\end{equation}
If the derived series terminates, i.e. there exists $k \in \N$ such that $\g^{(k)} = 0$, then $\g$ is a \emph{solvable Lie algebra}.

The \emph{lower central series}, which is of particular importance for our considerations in this paper,  $ \g = \g^{1} \supset \g^{2} \supset \ldots \supset \g^{k} \supset \ldots $ 
is again defined recursively
\begin{equation}
\g^{1}=\g, \qquad \g^{k} = [\g^{k-1},\g], \ k\geq 2. 
\end{equation}
If the lower central series terminates, i.e. there exists $k \in \N$ such that $\g^{k} = 0$, then $\g$ is 
called a \emph{nilpotent Lie algebra}. The largest value of $K$ for which we have $\g^{K} \neq 0$ is the \emph{degree of nilpotency} of the nilpotent Lie algebra $\g$. 

By definition, a nilpotent Lie algebra is also solvable. An Abelian Lie algebra is nilpotent of degree 1.

Because we have $ [\g^j,\g^k] \subset\g^{j+k} $ due to the Jacobi identity, the lower central series defines a \emph{natural filtration} on the Lie algebra $\g$.

The \emph{upper central series} is $ \z_{1} \subset\ldots \subset\z_{k} \subset\ldots \subset\g$ where $\z_1$ is the \emph{center} of $\g$ and $\z_k$ are defined recursively: $\z_k$ is the unique ideal in $\g$ such that $\z_k/\z_{k-1}$ is the center of $\g/\z_{k-1}$. The upper central series terminates, i.e. a number $L$ exists such that $\z_L=\g$, if and only if $\g$ is nilpotent \cite{Jac}.

We denote by $\cent(\h)$ the  \emph{centralizer} of a given subalgebra $\h \subset \g$ in $\g$
$$
\cent(\h) = \{ x \in \g | [x,y]=0, \ \forall y \in \h \}.
$$

We recall that an \emph{automorphism} $\Phi$ of a given Lie algebra $\g$ is a bijective linear map
$\Phi: \ \g \rightarrow \g $
such that for any pair $x,y$ of vectors in $\g$ 
\be\label{autom} 
\Phi([x,y])=[\Phi(x),\Phi(y)]. 
\ee
All automorphisms of $\g$ form a Lie group ${\rm Aut}(\g)$. Its Lie algebra is the algebra of \emph{derivations} of $\g$, i.e. of linear maps $D: \ \g \rightarrow \g $
such that for any pair $x,y$ of vectors in $\g$ 
\be\label{deriv} 
D([x,y])=[D(x),y]+[x,D(y)]. 
\ee
If a vector $z \in \g$ exists  such that 
$ D = {\rm ad} (z),$ i.e. $D(x)=[z,x], \ \forall x \in G,  $
the derivation $D$ is called an \emph{inner derivation}, any other one is an \emph{outer derivation}. The space of inner derivations is denoted $\inn (\g)$, of all derivations $\der(\g)$.

The ideals in the derived, lower and upper central series as well as their centralizers are invariant with respect to any automorphism and derivation, i.e. they belong among the \emph{characteristic ideals}.\medskip

We denote by $\dotplus$ the direct sum of vector spaces.

\section{Solvable Lie algebras with a given nilradical}\label{introsan}

Any solvable Lie algebra $\s$ contains a unique maximal nilpotent ideal $\n={\NR}(\s)$, the \emph{nilradical} of $\s$. We assume that $\n$ is known. That is, in some basis $( e_1, \ldots, e_n )$ of $\n$ we are given the Lie brackets
\be\label{nilkom}
[e_j,e_k] = {N_{jk}}^l e_l
\ee
(summation over repeated indices applies throughout the paper). 

Let us consider an extension of the nilpotent algebra $\n$ to a solvable Lie algebra $\s$, $\mathfrak{n} \subsetneq \mathfrak{s}$ having $\n$ as its nilradical. We call any such $\s$ a \emph{solvable extension} of the nilpotent Lie algebra $\n$. By definition, any such solvable extensions $\s$ is non--nilpotent. 

We can assume without loss of generality that the structure of $\s$ is expressed in terms of linearly independent vectors $f_1,\ldots,f_f\in \s$ added to the basis $( e_1, \ldots, e_n )$ so that together they form a basis of $\s$. The derived algebra of a solvable Lie algebra is contained in the nilradical (see \cite{Jac}), i.e.
\be\label{ssinn}
[\s,\s] \subset\n.
\ee
It follows that the Lie brackets in $\s$ take the form
\begin{eqnarray}\label{Agam1}
[f_a,e_j] & = & (A_a)^k_{j} e_k, \; 1 \leq a \leq f, \  1 \leq j \leq n, \\ 
\label{Agam2} [f_a,f_b] & = & \gamma^j_{ab} e_j, \; 1 \leq a<b \leq f.
\end{eqnarray}

The matrix elements of the matrices $A_a$ must satisfy certain linear relations following from the Jacobi identities $(f_a,e_j,e_k)$. The Jacobi identities $(f_a,f_b,e_j)$ provide 
linear expressions for the structure constants $\gamma^j_{ab}$ in terms of matrix elements of the commutators of  matrices $A_a$ and $A_b$. Finally, the Jacobi identities $(f_a,f_b,f_c)$ imply some bilinear compatibility conditions on $\gamma^j_{ab}$ and $A_a$ (which may become trivial for a particular choice of $\n$).

By inspection of Eq. (\ref{Agam1}) we realize that the matrices $A_a$ are matrices representing $f_a$ in the adjoint representation of $\s$ restricted to the nilradical $\n$, 
$$A_a = {\rm ad} (f_a)|_\n . $$

For any choice of $a$, the operator ${\rm ad} (f_a)|_\n$ is a derivation of $\n$. It must be an outer derivation -- if the contrary held then
$\n \dotplus {\rm span} \{ f_a \}$ 
would be a nilpotent ideal in $\s$ contradicting the maximality of the nilradical $\n$. In fact, the maximality of $\n$ implies that no nontrivial linear combination of the operators ${\rm ad} (f_a)|_\n$ can be a nilpotent matrix, i.e. ${\rm ad} (f_1)|_\n, \ldots, {\rm ad} (f_f)|_\n$ must be \emph{linearly nilindependent}. A nilpotent algebra $\n$ which possesses only nilpotent derivations and consequently is not a nilradical of any solvable Lie algebra is called \emph{characteristically nilpotent}.
 
In other words, finding all sets of matrices $A_a$ in Eq. (\ref{Agam1}) satisfying the Jacobi identity
is equivalent to finding all sets of outer nilindependent derivations of $\n$
\be
 D_1={\rm ad} (f_1)|_\n,\ldots,D_f={\rm ad} (f_f)|_\n.
\ee
Furthermore, in view of Eq. (\ref{ssinn}), the commutators $[D_a,D_b]$ must be inner derivations of $\n$. The structure constants $\gamma^j_{ab}$ in the Lie brackets (\ref{Agam2}) are determined through the consequence of Eq. (\ref{Agam2})
\be
[D_a,D_b] = \gamma^j_{ab} {\rm ad} ( e_j)|_\n
\ee
up to elements in the center $\z_1$ of $\n$. The consistency of $\gamma^j_{ab}$ and $D_i$ is then subject to the constraint
\be\label{fafbfc}
\gamma^j_{ab} D_c(e_j) +\gamma^j_{bc} D_a(e_j)+\gamma^j_{ca} D_b(e_j)=0
\ee
coming from the Jacobi identity $(f_a,f_b,f_c)$. We remark that l.h.s of Eq. (\ref{fafbfc}) is valued in the center of $\n$ because the derivations $D_a$ themselves satisfy the Jacobi identity. 

Different sets of derivations $D_a$ (and their accompanying constants $\gamma^j_{ab}$) may correspond to isomorphic Lie algebras. The equivalence between sets of derivations $D_a$ is generated by the following transformations:
\begin{enumerate}
\item We may add any inner derivation to any $D_a$.
\item We may simultaneously conjugate all $D_a$ by an automorphism $\Phi$ of $\n$, $D_a \rightarrow \Phi^{-1} \circ D_a \circ \Phi $.
\item We can change the basis in the space $ \sp \{ D_1,\ldots, D_f \}$.
\end{enumerate}
The corresponding changes in $\gamma^j_{ab}$ are of no interest to us in this paper so we don't explicitly write them out here.

\section{Upper bound on the dimension of a solvable extension of the given nilradical}\label{upperbound}

In the present section we derive the following upper bound on the maximal number $f$ of non--nilpotent elements $f_a$ we can add to a given nilradical $\n$.
\begin{theorem}\label{th1}
 Let $\n$ be a nilpotent Lie algebra and $\s$ a solvable Lie algebra with the nilradical $\n$. Let $\dim \n=n,\, \dim \s=n+f$. Then $f$ satisfies
\begin{equation}\label{mybound}
 f \leq \dim \n - \dim \n^2.
\end{equation}
\end{theorem}

In order to derive the estimate (\ref{mybound}) we start by choosing a convenient basis $\mathcal{E}$ of the nilpotent Lie algebra $\n$. 
We choose first some complement $\m_1$ of $\n^2$ in $\n$,
$$ \n = \n^2 \dotplus \m_1$$
 and denote $m_1=\dim \m_1$. We construct a
basis $\mathcal{E}_{\m_1}=(e_{n-m_1+1},\ldots,e_{n})$ of $\m_1$. In the next step, we recall that 
$$ \n^2 = [\n,\n] = [\m_1 \dotplus \n^2, \m_1 \dotplus \n^2 ] = [\m_1,\m_1] + \n^3$$
(the last sum is not necessarily direct). Consequently, we can choose a complement $\m_2$ of $\n^3$ in $\n^2$ such that $\m_2 \subset[\m_1,\m_1]$ and its basis $\mathcal{E}_{\m_2}$ in the form of some subset of Lie brackets of vectors in $\mathcal{E}_{\m_1}$, i.e.
$ \mathcal{E}_{\m_2}=(e_{n-m_1-m_2+1},\ldots,e_{n-m_1})$
where $m_2=\dim \m_2$ and for any $k\in \{ n-m_1-m_2+1, n-m_1\}$ exists a pair $y_k,z_k \in \mathcal{E}_{\m_1}$ (not necessarily unique) such that 
$ e_k = [y_k,z_k].$

Proceeding by induction we have
$$ \n^k = [\m_{k-1} \dotplus \n^k,\m_1 \dotplus \n^2] = [\m_{k-1},\m_1]+\n^{k+1} $$
and we can construct a complement $\m_k$ of $\n^{k+1}$ in $\n^k$, 
$$ \n^k = \n^{k+1} \dotplus \m_k,$$
$\m_k\subset[\m_{k-1},\m_1]$, $m_k=\dim \m_k$ and a basis
$ \mathcal{E}_{\m_k}=(e_{n+1-\sum_{i=1}^k m_i},\ldots,e_{n-\sum_{i=1}^{k-1} m_i})$ of $\m_k$ such that 
\begin{equation}\label{basiscomms}
 \forall e_j \in \mathcal{E}_{\m_k} \quad \exists y_j \in \mathcal{E}_{\m_{k-1}}, \, z_j \in \mathcal{E}_{\m_1} : \quad e_j=[y_j,z_j]. 
\end{equation} 
Together the elements of the bases $\mathcal{E}_{\m_k}$ form a basis $\mathcal{E}=(e_{1},\ldots,e_{n})$ of the whole nilpotent algebra $\n$. The main advantage of working in the basis $\mathcal{E}$ lies in the fact that any automorphism $\phi$, or any derivation $D$, is fully specified once its action on the elements of the basis $\mathcal{E}_{\m_1}$ of $\m_1$ is known due to the definition of an automorphism, Eq. (\ref{autom}), or a derivation, Eq. (\ref{deriv}), together with Eq. (\ref{basiscomms}), respectively. 

In particular this implies that the matrix of any automorphism $\Phi$ of $\n$ is upper block triangular in the basis $\mathcal{E}$ 
\begin{equation}
\Phi=\left( \begin{array}{cccc}
\Phi_{\m_K\m_K}& \ldots & \Phi_{\m_K\m_2} & \Phi_{\m_K\m_1}\\
 & \ddots &   & \vdots \\
&  & \Phi_{\m_2\m_2} & \Phi_{\m_2\m_1}\\
&  &  & \Phi_{\m_1\m_1}
\end{array}
 \right),
\end{equation} 
and its diagonal blocks $\Phi_{\m_k\m_k}, k=2,\ldots,K$, can be expressed as functions of the elements of the lowest diagonal block $\Phi_{\m_1\m_1}$ only by the repeated use of $\Phi([e_k,e_j])=[\Phi(e_k),\Phi(e_j)]$. The same applies to any derivation $D$ of $\n$, 
\begin{equation}
D=\left( \begin{array}{cccc}
D_{\m_K\m_K}& \ldots & D_{\m_K\m_2} & D_{\m_K\m_1}\\
 & \ddots & \vdots  & \vdots \\
&  & D_{\m_2\m_2} & D_{\m_2\m_1}\\
&  &  & D_{\m_1\m_1}
\end{array}
 \right).
\end{equation} 
Due to the relation $D([e_k,e_j])=[D(e_k),e_j]+[e_k,D(e_j)]$ we conclude that elements of the diagonal blocks $D_{\m_k\m_k}, k=2,\ldots,K$ are linear functions of elements of $D_{\m_1\m_1}$. 
E.g. for $e_j\in \m_2$, $e_j=[e_k,e_l]$ where $e_k,e_l \in \m_1$ we have
\begin{eqnarray*}
 D(e_j)  & = & \sum_{p=n-m_1+1}^{n} \left( {D^{p}}_{k} [e_p,e_l] + {D^{p}}_{l}[e_k,e_p]\right) \mod \n^3\\
& = & \sum_{p=n-m_1+1}^{n} \sum_{q=n-m_1-m_2+1}^{n-m_1} \left( {D^{p}}_{k} {N_{pl}}^q + {D^{p}}_{l} {N_{kp}}^q \right) e_q \mod \n^3.
\end{eqnarray*}
i.e.
$${D^{q}}_{j}= \sum_{p=n-m_1+1}^{n} \left( {D^{p}}_{k} {N_{pl}}^q + {D^{p}}_{l} {N_{kp}}^q \right).$$
showing that any $D_{\m_2\m_2}$--block element ${D^{q}}_{j}$ can be expressed in terms of $D_{\m_1\m_1}$--block elements together with the structure constants ${N_{ab}}^c$ of $\n$ and that its dependence on $D_{\m_1\m_1}$ is a linear one.

Extending the same argument, we see that in general $D_{\m_j\m_k}, k\leq j=2,\ldots,K$ is a linear function of elements in the last column blocks $D_{\m_1\m_1},\ldots,D_{\m_{j-k+1}\m_1}$.

We notice that inner derivations have strictly upper triangular block structure because inner derivations by definition map $\n^k\rightarrow \n^{k+1}$. Consequently, any set of outer derivations $\{ D_1,\ldots,D_f \} $ such that $[D_j,D_k]\in \inn(\n)$  for all $j,k=1,\ldots,f$ must necessarily have commuting $\m_1\m_1$--submatrices,
\begin{equation}\label{d11comms}
[(D_j)_{\m_1\m_1},(D_k)_{\m_1\m_1}]=0.
\end{equation}
A derivation $D$ is nilpotent if and only if its submatrix $D_{\m_1\m_1}$ is nilpotent. One of these implications is obvious; the other is derived as follows. From a consequence of Eq. (\ref{deriv})
$$ D^n[x,y] = \sum_{j=0}^n {{n}\choose{j}}  [D^j x, D^{n-j}y]$$
together with the assumed existence of $N\in \N$ such that $(D_{\m_1\m_1})^N=0$ we deduce the existence of $M\in \N$ such that $(D_{\m_k\m_k})^M=0$ for all $k=1,\ldots,K$, i.e. the block upper triangular matrix of $D^M$ has vanishing diagonal blocks and is consequently nilpotent, implying also the nilpotency of $D$ itself.

This equivalence implies that derivations $D_1,\ldots,D_f$ are linearly nilindependent if and only if their submatrices $(D_1)_{\m_1\m_1},\ldots,(D_k)_{\m_1\m_1}$ are linearly independent. Together with Eq. (\ref{d11comms}) this means that the number $f$ of linearly nilindependent outer derivations of the given nilpotent algebra $\n$ commuting to inner derivations is bounded from above by the maximal number of linearly independent commuting matrices of dimension $m_1\times m_1$. This number is $m_1=\dim \n - \dim \n^2$, finishing the derivation of the estimate (\ref{mybound}). $\square$
\medskip

\medskip

The estimate (\ref{mybound}) allows us to construct also a lower bound on dimension of the nilradical of a given solvable Lie algebra $\s$. We have
$$ \dim \s +\dim \n^2 \leq 2 \dim \n $$
and $\s^{(2)}=(\s^2)^2 \subset\n^2$ because $\s^2 \subset\n$. Together we find
\begin{equation}\label{boundonn}
 \dim \n \geq \frac{1}{2} \left( \dim \s + \dim \s^{(2)} \right).
\end{equation}
In our experience, this estimate is often less accurate than the trivial estimate $\dim \n \geq \dim \s^2$. Nevertheless, the bound (\ref{boundonn}) can be useful in some particular cases.

\section{Levi decomposable algebras with the given nilradical}\label{levinilr}

Let us now move our attention to nonsolvable algebras with the given nilradical $\n$. As was demonstrated by Levi \cite{Levi}, any such algebra $\g$ can be written in the form
\begin{equation}\label{Levidec}
\g=\r\dotplus \p, \; \r \supset \n, \;[\r,\g]\subset \n, [\p,\p]=\p 
\end{equation}
where $\r$ is the \emph{radical}, i.e. the maximal solvable ideal of $\g$, and $\p$ is a semisimple Lie algebra, called the \emph{Levi factor}, unique up to automorphisms of $\g$. We recall that $\ad (\p) |_{\r}$ provides a representation of $\p$ on $\r$ and this fact is a cornerstone in the construction and classification of algebras of this type. We shall consider only the case when $\g$ is \emph{indecomposable} in the sense that it cannot be decomposed into a direct sum of ideals. This assumption in particular implies that the representation  $\ad (\p)|_{\r}$ of $\p$ on $\r$ is faithful, i.e. a monomorphism into $\der (\r)$.

The classification of algebras of the type (\ref{Levidec}) with $\p\neq 0,\r \neq 0$, i.e. of \emph{Levi decomposable} algebras, was considered in \cite{Tur1} and \cite{Tur4}. The approach used there was to consider a given semisimple algebra $\p$ and all its possible representations $\rho$ on a vector space $V$ of chosen dimension. For each $\rho$, all solvable algebras $\r$ compatible with the representation $\rho$ were found by an explicit evaluation of the Jacobi identity with unknown structure constants ${c_{ij}}^{k}$ of the radical $\r$, and classified into equivalence classes. 

In \cite{Tur4} also some general properties of Levi decomposable algebras were found and used in the construction of all 9--dimensional Levi decomposable algebras\footnote{with one additional algebra missing, as was shown in \cite{RCSmissingalg}}. These properties are a direct consequence of the complete reducibility of representations of semisimple Lie algebras. Namely, 
\begin{enumerate}
\item if the representation $\ad (\p)|_{\r}$ of $\p$ is irreducible then $\r$ is Abelian.
\item If $\r$ is solvable non--nilpotent, then there exists a complement $\q$ of $\n$ in $\r$, i.e.
$$ \r = \n \dotplus \q$$
such that $\ad  (p)|_{\q}=0$ for all $p\in\p$, i.e. $\ad (\p)|_{\r}$ must necessarily contain a copy of the trivial representation.

In view of this property, it is of interest to study and classify Levi decomposable algebras with nilpotent radicals first.
\item The set of all elements belonging to the trivial representation
$$ \{ x\in\n | [p,x]=0, \, \forall p \in \p \} $$
is a subalgebra of $\n$.
\end{enumerate}

In this Section we intend to provide several more stringent, yet easy to verify, restrictions on the structure of $\n$ obtained from the compatibility of the nilradical structure with the given representation of $\p$. We shall call any Levi decomposable algebra $\g$ with the nilradical $\n$ (radical $\r$) a \emph{Levi extension} of $\n$ ($\r$, respectively). In most of this Section we shall suppose that the nilradical coincides with the radical and that the Levi factor acts faithfully on $\n$.  

Because all ideals in the characteristic series and their centralizers are invariant with respect to all derivations, in particular with respect to $\ad (\p)|_{\r} $, we can use Lie's theorem to easily deduce the following proposition. 

\begin{prop}\label{ldbasiccrit}
If a complete flag 
$$ 0 \subsetneq  V_1 \subsetneq  V_2 \subsetneq \ldots \subsetneq  V_{n} = \n   $$
of codimension 1 subspaces can be built out of ideals in the characteristic series and their centralizers,
then no Levi decomposable algebra $$\g=\n \dotplus \p$$ such that $[\p,\n]\neq 0$ exists. 
\end{prop}

Using this criterion one can immediately establish, without further considerations of the structure of the representations of $\p$, that out of low dimensional nilpotent algebras (dimension at most 5), the following can never appear as a nilradical of a Levi decomposable algebra\footnote{Our notation follows \cite{PSW} for the nilpotent algebras and \cite{Tur1,Tur4} for their Levi extensions.}
\begin{itemize}
\item $\dim \n=4:$
\begin{description}
\item{$A_{4,1}$:} $[e_2,e_4]=e_1,\,[e_3,e_4]=e_2$; the characteristic flag is 
$$0 \subset \n^3 \subset \n^2 \subset \cent (\n^2)\subset \n.$$ 
\end{description}
\item $\dim \n=5:$
\begin{description}
\item{$A_{5,2}$:} $[e_2,e_5]=e_1,\,[e_3,e_5]=e_2,\,[e_4,e_5]=e_3$; the characteristic flag is $$0 \subset \n^4 \subset \n^3 \subset \n^2 \subset \cent (\n^3)\subset \n.$$
\item{$A_{5,5}$:} $[e_3,e_4]=e_1,\,[e_2,e_5]=e_1,\,[e_3,e_5]=e_2$; the characteristic flag is $$0 \subset \n^3 \subset \n^2 \subset \z_2 \subset \cent (\n^2)\subset \n.$$ 
\item{$A_{5,6}$:} $[e_2,e_5]=e_1,\,[e_3,e_4]=e_1,\,[e_3,e_5]=e_2,\,[e_4,e_5]=e_3$; with the same flag as $A_{5,2}$.
\end{description}
\end{itemize}
It is rather interesting that all other indecomposable nilpotent algebras $A_{3,1}$, $A_{5,1},A_{5,3},A_{5,4}$ of dimension $2\leq n \leq 5$ do show up as nilradicals in the Turkowski list of Levi decomposable algebras of dimension $\leq 8$, i.e. they do admit Levi extension(s). In the case of six--dimensional nilpotent algebras, the same argument allows to immediately exclude from the list of Levi extendable nilradicals the algebras $A_{6,1},A_{6,2},A_{6,6},A_{6,7},A_{6,11},A_{6,16},A_{6,17},A_{6,19},A_{6,20},A_{6,21},A_{6,22}$. In this case, however, not all of the remaining algebras allow a Levi extension, as a brief look into \cite{Tur4} tells us. According to Turkowski, only four algebras $A_{6,3},A_{6,4},A_{6,5},A_{6,12}$ out of 22 indecomposable 6--dimensional nilpotent algebras contained in the list in \cite{PSW} allow a Levi extension. 

We can improve on Proposition \ref{ldbasiccrit} by considering the following type of ideals:
let $\mathfrak{i},\mathfrak{j}$ be characteristic ideals in a Lie algebra $\g$ and let
$ \mathfrak{k}= \{ x\in\g | [x,y]\in\mathfrak{j}, \; \forall y\in \mathfrak{i}\} $
be a subspace in $\g$. Then $ \mathfrak{k}$ is a characteristic ideal because
$$ [Dx,y]=D[x,y]-[x,Dy]\in \mathfrak{j}$$  
by the virtue of the definition of $\mathfrak{k}$ and the characteristic property of $\mathfrak{i}, \mathfrak{j}$, i.e. $Dy\in \mathfrak{i}$, $D[x,y]\in \mathfrak{j}$; consequently, $ \mathfrak{k}$ is closed under every derivation, i.e. is itself characteristic. Such ideals, if identified, can be used to refine the sequence of characteristic subspaces in Proposition \ref{ldbasiccrit}. 

Concerning the decomposable nilpotent algebras, we notice that e.g. $A_{1,1}+A_{3,1}$, i.e. centrally extended Heisenberg algebra, appears in \cite{Tur1} only as a nilradical of a Levi extendable 5--dimensional solvable radical, but not a nilpotent radical of a 7--dimensional Levi decomposable algebra. The reason is that the Levi extension $\s\l(2) \dotplus (A_{1,1}+A_{3,1})$ is decomposable. Similarly also for some other decomposable nilpotent algebras.
\medskip

Now we employ an analysis similar to the one in Section \ref{upperbound} in order to present a more refined necessary criteria on the interplay between the representation $\ad(\p)|_{\n}$ of the semisimple algebra $\p$ on the nilpotent radical $\n$ and the structure of $\n$. 

The complete reducibility of representations of semisimple Lie algebras allows us to deduce the existence of complementary $\ad(\p)$--invariant subspaces $\tilde{\m}_j$ of $\n^{j+1}$ in $\n^j$
\begin{equation}
\n^j=\tilde{\m}_j\dotplus \n^{j+1}, \; \ad (\p) \tilde{\m}_j \subset \tilde{\m}_j, \; j =1,\ldots,K. 
\end{equation}
Let us now explore whether these subspaces can be taken in the same form as the one used in Section \ref{upperbound} (or as close to it as possible). We can take 
$$ \m_1 = \tilde\m_1.$$
Now the commutator of two $\ad (\p)$--invariant subspaces is again an $\ad (\p)$--invariant subspace (see the definition of a derivation, Eq. (\ref{deriv})). In particular, $[\m_1,\m_1]$ is an $\ad (\p)$--invariant subspace
of $\n^2$ and we have
$$ \n^2 = [\m_1,\m_1] + \n^3.$$
Since both $[\m_1,\m_1]$ and $\n^3$ are $\ad (\p)$--invariant, so is their intersection $[\m_1,\m_1] \cap \n^3$. By the complete reducibility of $\ad (\p)$, there is an $\ad (\p)$--invariant complement of $[\m_1,\m_1] \cap \n^3$ in $[\m_1,\m_1]$ which we denote $\m_2$. Altogether, we have
$$
\n^2 = \m_2 \dotplus \n^3, \; \m_2\subset [\m_1,\m_1], \; \ad (\s) \m_2 \subset \m_2. 
$$
Continuing in the same way, we can construct a sequence of subspaces $\m_j$ such that
\begin{equation}\label{nintoms}
 \n= \m_K \dotplus \m_{K-1} \dotplus \ldots \dotplus \m_1 
\end{equation}
where 
\begin{equation}
 \n^j= \m_j \dotplus \n^{j+1}, \; \m_j\subset [\m_{j-1},\m_1], \; \ad (\p) \m_j \subset \m_j.
\end{equation}
The only minor difference between the decomposition constructed here and in Section \ref{upperbound} is that now we cannot in general find a basis of $\m_1$ such that bases of $\m_j$ are obtained by simple commutations (cf. Eq. (\ref{basiscomms})) -- in the present case taking linear combinations of the commutators may be necessary.
Nevertheless, all the essential arguments presented in Section \ref{upperbound} can be taken over here.

We have established that in any basis of the nilradical $\n$ which respects the decomposition (\ref{nintoms}) the matrices of $\ad(\p)|_{\n}$ have block diagonal form. If any of the blocks is 1--dimensional then it necessarily corresponds to the trivial representation $\rho(p)=0, \, \forall p\in\p$. Similarly as in Section \ref{upperbound}, the $\m_j\m_j$--submatrices of the representation $\ad(\p)|_{\n}$, $j>1$, i.e. the matrices of $\ad(\p)|_{\m_j}$, are fully determined by $\ad(\p)|_{\m_1}$
through linear relations coming from the definition of a derivation Eq. (\ref{deriv}).

For the same reason, the kernel of $\ad(\p)|_{\m_1}$ is also the kernel of $\ad(\p)|_{\n}$. Therefore, the representation $\ad(\p)|_{\n}$ of the Levi factor $\p$ is faithful if and only if   $\ad(\p)|_{\m_1}$ is faithful.

In the particular case when $\m_j$ is one--dimensional we can easily find a simple relation between the representations $\ad(\p)|_{\m_1}$ and $\ad(\p)|_{\m_{j+1}}$. Let $\m_{j}=\span \{ x \}$. 
We have
\begin{equation}\label{adsxy}
 \ad(p) [x,y]= [x, \ad(p) y], \forall y\in \m_1, \, \forall p\in\p.
\end{equation}
Now let us assume that $y$ belongs to a nontrivial irreducible representation $\rho$ contained in $\ad(\p)|_{\m_1}$ and that $[x,y]\neq 0$. Let $V$ be the space generated by repeated applications of $\ad(p),p\in\p$ on $y$, i.e. $V$ is the representation space of the representation $\rho$. Consider $\ad(x)V \subset [\m_{j},\m_{1}]$. Due to Eq. (\ref{adsxy}) the subspace $\ad(x)V$ is by construction invariant with respect to $\ad(\p)$. The kernel of $\ad(x):V\rightarrow \ad(x)V$ is an invariant subspace of $V$. By assumption the representation $\rho$ is irreducible and $\ad(x)V\neq 0$. Therefore, $\ad(x)V$ is a vector space isomorphic to $V$ and an irreducible representation $\rho'$ equivalent to $\rho$ is contained in the decomposition of the representation $\ad(\p)|_{[\m_j,\m_1]}$ into irreducible representations. 

To sum up, let $\rho_1\oplus\rho_2\oplus\ldots\oplus\rho_L$ be a decomposition of $\ad(\p)|_{\m_1}$ into irreducible representations. Then the induced representation on $[\m_j,\m_1]$ is a direct sum of irreducible representations equivalent to the ones contained in some subset $\left\{ \rho_{a}, a\in J\subset \{1,\dots,L \} \right\}$.
The same necessarily holds also for $\m_{j+1}\subset [\m_j,\m_1]$.

If $\dim\m_j>1$ then the relation between the representations on $\m_1,\m_j,\m_{j+1}$ takes a more complicated form. Because the commutators of $e_a\in \m_j,e_b\in\m_1$ transform under action of any block--diagonal derivation $D$ by
\begin{equation}\label{Deaeb}
{D {[e_a,e_b]}}= \sum_{c=n+1-\sum_{i=1}^j m_i}^{n-\sum_{i=1}^{j-1} m_i} {D^c}_{a} [e_c,e_b]+ \sum_{d=n+1-m_1}^{n} {D^d}_{b} [e_a,e_d]
\end{equation}
where ${D^c}_{a}$ are components of the matrix of $D|_{\m_j}:\m_j\rightarrow \m_j$ and ${D^d}_{b}$ are components of $D|_{\m_1}:\m_1\rightarrow \m_1$, the commutator subspace $[\m_j,\m_1]$ transforms under a certain subset\footnote{It may be only a subset because some of the commutators in Eq. (\ref{Deaeb}) may vanish.} of irreducible factors in the tensor representation $\ad(\p)|_{\m_j} \otimes \ad(\p)|_{\m_1}$ and the same is true also for $\m_{j+1}\subset [\m_j,\m_1]$.

We sum up these conclusions in the following theorem 
\begin{theorem}\label{strLevnilp}
Let $\g$ be an indecomposable Lie algebra with a nilpotent radical $\n$ and a nontrivial Levi decomposition
$$ \g= \n \dotplus \p.$$
There exists a decomposition of $\n$ into a direct sum of $\ad(\p)$--invariant subspaces
$$
 \n= \m_K \dotplus \m_{K-1} \dotplus \ldots \dotplus \m_1 
$$
where
$$
 \n^j= \m_j \dotplus \n^{j+1}, \; \m_j\subset [\m_{j-1},\m_1], \; \ad (\p) \m_j \subset \m_j.
$$
such that
$\ad(\p)|_{\m_1}$ is a faithful representation of $\p$ on $\m_1$.  For $j=2,\ldots,K$ the representation $\ad(\p)|_{\m_{j}}$ of $\p$ on the subspace $\m_j$  can be decomposed into some subset of irreducible components of the tensor representation $\ad(\p)|_{\m_{j-1}} \otimes \ad(\p)|_{\m_1}$.

If any of the subspaces $\m_j$ is one--dimensional then $\ad(\p)|_{\n}$ must contain a copy of the trivial representation corresponding to the subspace $\m_j$. When $j<K$, the representation of $\p$ on 
$\m_{j+1}$ can be decomposed into a sum of irreducible representations, each of which is equivalent to an irreducible representation contained in the decomposition of $\m_1$. 

In particular, when  $\ad(\p)|_{\m_1}$ is irreducible and $\dim \m_j=1, \, 1<j<K$ then the representation $\ad(\p)|_{\m_{j+1}}$ on $\m_{j+1}$ is equivalent to $\ad(\p)|_{\m_1}$. 
\end{theorem}
We remark that it was shown in \cite{RCSnsla} that when the radical $\r$ of a Levi decomposable algebra $\g$ has a one--dimensional center then the representation $\ad(\p)|_{\n}$ contains a copy of the trivial representation. This result is contained in our theorem as a particular subcase when $\dim \m_K=1$.

Theorem \ref{strLevnilp} gives us a simple dimensional necessary criterion on possible Levi extensions of $\n$. Namely, a faithful representation of dimension $m_1=\dim \n - \dim \n^2$ must exist. For example, there cannot be any Levi extension of the Heisenberg algebra $A_{3,1} \; ([e_2,e_3]=e_1)$ with a Levi factor other than $\sl(2)$ and similarly for any other nilradical such that $\dim \n - \dim \n^2=2$. 

In particular, if $\n$ is \emph{filiform}, i.e. a nilpotent Lie algebra of maximal degree of nilpotency, $K=n-1$ \cite{Vergne,EchGN,GKh1,GKh2}, we have $m_1=\dim \n - \dim \n^2=2$ and $m_j=\dim \n^j-\dim \n^{j+1}=1$ for $j=2,\ldots,n-1$. When $\dim\n=3$ we have the Heisenberg algebra which possesses a Levi extension. When $\dim\n\geq 4$ the existence of a Levi extension would imply that the 1--dimensional subspace $\m_3$ must carry an equivalent copy of the 2--dimensional irreducible representation of $\sl(2)$ on $\m_1$, i.e. a clear contradiction. Therefore, no Levi decomposable algebra with a filiform nilradical $\n$ ($\dim\n\geq 4$) exists as was already derived in \cite{GKh2} -- Lemma 25 and independently in \cite{ABCSGV} -- Corollary 1 by other means.

In the same way we can also explain the prevalence of Levi factors isomorphic to $\sl(2)$ in Turkowski's lists of real Levi decomposable algebras. Whenever we can identify a nontrivial 2--dimensional representation in the subspace $\m_1$ the Levi factor $\so(3)$ is immediately ruled out, e.g. for the nilpotent algebra ${\bf A_{5,3}}$ with the dimensions of the invariant subspaces $m_1=2,\, m_2=1,\, m_3=2$. Even when $\m_1$ can carry a three--dimensional irreducible representation there may be further restrictions. They come from the fact that for $\so(3)$ we have ${\bf 3} \otimes {\bf 3}={\bf 5} \dotplus {\bf 3} \dotplus {\bf 1}$ where ${\bf 5} \dotplus {\bf 1}$ is the symmetric part and ${\bf 3}$ the antisymmetric part. Because $[e_i,e_j]$ is antisymmetric, only  ${\bf 3}$ remains in $[V,V]$ where $V={\bf 3}$. Consequently, the algebra 
$${\bf A_{5,1}:}\; [e_3,e_5]=e_1,\,[e_4,e_5]=e_2$$
with $\n^2=\z_1=\span\{e_1,e_2\}$ cannot have a Levi extension with the Levi factor $\so(3)$. If the contrary held we would have $\m_1={\bf 3}, \m_2={\bf 1}\dotplus {\bf 1}$ but ${\bf 1}\dotplus {\bf 1}$ is not contained in $[{\bf 3},{\bf 3}]\simeq {\bf 3}$ of $[\m_1,\m_1]$.

Let us now apply these ideas to 6--dimensional nilpotent radicals. Let us consider the nilpotent algebra
$$ {\bf A_{6,15}:} \, [e_1,e_2]=e_3+e_5,\,[e_1,e_3]=e_4,\,[e_1,e_4]=e_6, [e_2,e_5]=e_6$$
which does not appear in Turkowski's list of Levi decomposable algebras as a radical. 
It has an incomplete flag of characteristic ideals 
$$0 \subset \n^4 \subset \n^3 \subset \n^2 \subset \z_3 \subset \n$$
in which only a 5--dimensional ideal is missing. Therefore, if any Levi decomposable algebra with the radical ${\bf A_{6,15}}$ exists then the action of the Levi factor $\s$ on the 4--dimensional ideal $\z_3$ is trivial as is seen by dimensional analysis and we have a decomposition of the subspaces $\m_i$ into irreducible representations as follows
$$ \m_1= {\bf 2}_1 \dotplus {\bf 1}_1, \;  \m_2= {\bf 1}_2, \; \m_3= {\bf 1}_3,  \; \m_4= {\bf 1}_4$$
where the boldface numbers stand for representation spaces of irreducible representations of $\s$ of that dimension and indices specify into which $\m_i$ space they belong. The representation space ${\bf 1}_1$ coincides with the 1--dimensional subspace $\m_1\cap \z_3$.
By Theorem \ref{strLevnilp} we have 
\begin{equation}\label{m3m4}
 \m_3= [{\bf 1}_1,\m_2]=[\m_1\cap \z_3,\m_2], \; \m_4= [{\bf 1}_1,\m_3]=[\m_1\cap \z_3,\m_3]. 
\end{equation}
At the same time, $\z_3$ of the algebra ${\bf A_{6,15}}$ is Abelian and contains both $\m_2$ and $\m_3$ which leads to a contradiction with Eq. (\ref{m3m4}). Therefore, no Levi extension of ${\bf A_{6,15}}$ exists. The same argument can be applied also to the algebra ${\bf A_{6,17}}$.

The case of algebras ${\bf A_{6,8},A_{6,9}}$ is more involved. They can be both viewed as an extension of the 5--dimensional algebra ${\bf A_{5,3}}$ 
$$[e_3,e_4]=e_2,\,[e_3,e_5]=e_1,\,[e_4,e_5]=e_3$$
by one element $e_6$ which has only non--vanishing commutators with $e_4,e_5$, spanning a one--dimensional subspace in the center $\z_1=\span\{e_1,e_2\}$. In a suitable basis we have
$ [e_6,e_4]=e_2$ in ${\bf A_{6,8}}$ and $ [e_6,e_4]=e_1$ in ${\bf A_{6,9}}$, respectively. Whereas the 5--dimensional algebra ${\bf A_{5,3}}$ does possess a Levi extension by $\sl(2)$, neither ${\bf A_{6,8}}$ nor ${\bf A_{6,9}}$ do. The reason is that the additional element $e_6$ presents an obstruction which can be identified in the following way.
We have $\n^2=\span\{e_1,e_2,e_3\}, \, \n^3=\z_1=\span\{e_1,e_2\}$, $\z_2= \span\{e_1,e_2,e_3,e_6\}$.
By dimensional analysis we find that the only hypothetically permissible Levi extension is by the factor $\p=\sl(2)$ and the structure of the representations must be as follows:
$$ \m_1={\bf 2_1} \oplus {\bf 1_1}, \; \m_2={\bf 1_2}, \; \m_3={\bf 2_3}. $$
where ${\bf 1_1} \subset \z_2$. The basis respecting the characteristic subspaces can be chosen without loss of generality in the form
\begin{eqnarray*}
\tilde e_1 &  =  & e_1, \quad \tilde e_2 = e_2, \quad n^3=\m_3={\bf 2_3}=\span \{ e_1,e_2 \}, \\
\tilde e_3 &  =  & e_3 \mod \n^3, \quad \m_2={\bf 1_2}=\span \{ \tilde e_3 \}, \\
\tilde e_6 &  =  & e_6 \mod \n^2, \quad {\bf 1_1}=\span \{ \tilde e_6 \}, \\
\tilde e_4 &  =  & e_4 \mod \z_2, \quad \tilde e_5  =   e_5 \mod \z_2, \quad {\bf 2_1}=\span \{ \tilde e_4, \tilde e_5 \},
\end{eqnarray*}
where e.g. $\mod \n^2$ stands for some (unknown) element of $\n^2$. Because both $\tilde e_6$ and $\tilde e_3$ belong to the trivial representation of $\sl(2)$, we can add a suitable multiple of $\tilde e_3$ to $\tilde e_6$ 
to set
$$ \tilde e_6 = e_6 \mod \n^3$$
without altering the block diagonal structure of $\ad(\p)$ acting on $\m_1\dotplus \m_2\dotplus \m_3$.
Now we have ${\bf 1_1}=\span \{\tilde e_6 \}$ and $[{\bf 1_1},{\bf 2_1}]\subset [{\bf 1_1},\n]=V$ where $V$ is a certain 1--dimensional subspace of the center $\m_3$ ($V=\span\{ e_2 \}$ for ${\bf A_{6,8}}$, $V=\span\{ e_1 \}$ for ${\bf A_{6,9}}$). That means we have arrived at a contradictory conclusion that a 1--dimensional subspace must carry a 2--dimensional representation and consequently no Levi extension of algebras ${\bf A_{6,8}}$ and ${\bf A_{6,9}}$ exists. The same holds also for ${\bf A_{6,10}}$ which is just another real form of the complex version of ${\bf A_{6,8}}$.

A similar but somewhat simpler argument shows that ${\bf A_{6,13}}$ 
\begin{equation}\label{A613}
[e_1,e_2]=e_5, \, [e_1,e_3]=e_4, \, [e_1,e_4]=e_6, \, [e_2,e_5]=e_6
\end{equation}
does not possess a nontrivial Levi decomposition.
Namely, we have 
$$ \n^2=\z_2=\span\{e_4,e_5,e_6\}, \n^3=\z_1=\span\{e_6\}, \cent (\n^2)=\span\{ e_3,e_4,e_5,e_6 \}.$$
By dimensional analysis alone we have the following structure of irreducible representations of hypothetical $\ad(\p)$: 
$$ \m_1= {\bf 2_1} \dotplus {\bf 1_1}, \;  \m_3= {\bf 1_3} $$
and two options for $\m_2$: either $\m_2={\bf 2_2}$ or $\m_2={\bf 1}\dotplus {\bf 1}$.
Out of the two, $\m_2={\bf 1}\dotplus {\bf 1}$ cannot be found in the antisymmetrized tensor product of  ${\bf 2_1}\dotplus {\bf 1_1}$ with itself; therefore, it must be $\m_2={\bf 2_2}=[{\bf 2_1},{\bf 1_1}]$. On the other hand,
from the Lie brackets (\ref{A613}) we have
$$ [{\bf 2_1},{\bf 1_1}] \subset [\cent (\n^2),\n] = \span\{ e_4,e_6\}$$
which splits into $\m^3$ and a 1--dimensional subspace of $\m^2$. Therefore, $[{\bf 2_1},{\bf 1_1}]$ must be simultaneously a trivial representation and 2--dimensional irreducible representation, a contradiction showing that no Levi extension of ${\bf A_{6,13}}$  exists.

That leaves only the nilpotent algebras ${\bf A_{6,14}}$ 
$$ [e_1,e_3]=e_4, \; [e_1,e_4]=e_6, \; [e_2,e_3]=e_5, \; [e_2,e_5]= \epsilon e_6, \; \epsilon^2=1 $$
and ${\bf A_{6,18}}$
$$ [e_1,e_2]=e_3, \; [e_1,e_3]=e_4, \; [e_1,e_4]=e_6, \; [e_2,e_3]= e_5, \; [e_2,e_4]= e_6 $$
unexplained. 
${\bf A_{6,14}}$ has the characteristic ideals
$$ \n^2=\z_2=\span\{ e_4,e_5,e_6 \}, \n^3=\z_1=\span \{ e_6 \}, \; \cent (\n^2) =\span\{ e_3,e_4,e_5,e_6 \} $$
which seem to allow the representation structure of the Levi factor in the form
$$ \m_1={\bf 2_1}\dotplus {\bf 1_1}, \; \m_2={\bf 2_2}, \;  \m_3={\bf 1_3} $$
with $\cent \n^2={\bf 1_1}\dotplus \m_2\dotplus \m_3$. We did not find any obvious obstruction to this representation structure considering dimension only. Therefore, in order to exclude the possible existence of a Levi extension in this case we have to consider the representation $\ad(\p)|_\n$ in more detail. We can assume that
\begin{eqnarray*}
{\bf 2_1} & = & \span\{ \tilde e_1= e_1\mod \cent(\n^2), \tilde e_2= e_2\mod \cent(\n^2)\}, \\
{\bf 1_1} & = & \span\{ \tilde e_3= e_3\mod \n^2\}, \\
{\bf 2_2} & = & \span\{ \tilde e_4= e_4\mod \n^3, \tilde e_5= e_5\mod \n^3 \}, \\
{\bf 1_2} & = & \span\{ \tilde e_6= e_6\}
\end{eqnarray*}
and consider $\p=\left(\begin{array}{cc} 1 & 0 \\ 0 & -1 \end{array} \right)\in\sl(2)$ represented in this representation space. We have by assumption
$$ \ad(p) \tilde e_1=\tilde e_1,\;\ad(p) \tilde e_2=-\tilde e_2, \; \ad(p) \tilde e_3=0$$
together with its consequences
$$ \ad(p) \tilde e_4= \ad(p) [\tilde e_1,\tilde e_3]=\tilde e_4,\;\ad(p) \tilde e_5= \ad(p) [\tilde e_2,\tilde e_3]=-\tilde e_5$$
and finally
$$ \ad(p) \tilde e_6= \ad(p) [\tilde e_1,\tilde e_4]= 2 \tilde e_6 $$
in a clear violation of $$ \m_3=\span\{ \tilde e_6\}= {\bf 1_3} .$$

Similarly for ${\bf A_{6,18}}$ where the characteristic ideals are 
$$ \n^2=\z_3=\span\{ e_3,e_4,e_5,e_6 \}, \n^3=\z_2=\span \{ e_4,e_5,e_6 \},$$
and $\n^4=\z_1=\span \{ e_6 \} $, i.e. a similar representation structure
$$ \m_1={\bf 2_1}, \; \m_2={\bf 1_2}, \; \m_2={\bf 2_3}, \;  \m_4={\bf 1_4} $$
naively seems possible. Now we have
\begin{eqnarray*}
\m_1 & = & \span\{ \tilde e_1= e_1\mod \n^2), \tilde e_2= e_2\mod \n^2\}, \\
\m_2 & = & \span\{ \tilde e_3= e_3\mod \n^3\}, \\
\m_3 & = & \span\{ \tilde e_4= e_4\mod \n^4,  \tilde e_5= e_5\mod \n^4 \}, \\
\m_4 & = & \span\{ \tilde e_6= e_6\}.
\end{eqnarray*}
If the representation of $\sl(2)$ on $\n$ exists we can again consider $\p=\left(\begin{array}{cc} 1 & 0 \\ 0 & -1 \end{array} \right)\in\sl(2)$ acting on $\n$. We have again
$$ \ad(p) \tilde e_1=\tilde e_1,\;\ad(p) \tilde e_2=-\tilde e_2$$
together with its consequences 
$$ \ad(p) \tilde e_3=\ad(p)[\tilde e_1,\tilde e_2]=0, \; \ad(p) \tilde e_4= \ad(p) [\tilde e_1,\tilde e_3]=\tilde e_4,\;\ad(p) \tilde e_5= \ad(p) [\tilde e_2,\tilde e_3]=-\tilde e_5$$
and
$$ \ad(p) \tilde e_6= \ad(p) [\tilde e_1,\tilde e_4]= 2 \tilde e_6 $$
demonstrating the incompatibility of $\sl(2)$ with ${\bf A_{6,18}}$.

Concerning the $\so(3)$ Levi factor acting on 6--dimensional indecomposable nilpotent radicals, we may consider only the four algebras which have not be excluded by the previous analysis (and appear as nilpotent radicals in Turkowski's list \cite{Tur4}). We can establish on dimensional grounds that ${\bf A_{6,12}}$ with the flag $0\subsetneq \n^3\subsetneq \n^2 \subsetneq \cent (\n^2) \subsetneq \n$ of dimensions $(0,1,2,4,6)$ cannot have a Levi extension by  $\so(3)$. ${\bf A_{6,3}}$ with dimensions $m_1=m_2=3$ is the only one to allow a three--dimensional representation of $\so(3)$ (in fact two copies of it) in its Levi extension, identified as  $L_{9,11}$. Out of the remaining two Levi extendable algebras ${\bf A_{6,4}}$, ${\bf A_{6,5}}$ with $m_1=4,\, m_2=2$ only the second one has a Levi extension $L_{9,4}$ with the Levi factor $\so(3)$ (with the 4--dimensional bispinor representation of $\so(3)$ acting on $\m_1$ and the trivial representation on $\m_2$). The fact that ${\bf A_{6,4}}$ does not have a nontrivial Levi extension by $\so(3)$ cannot be found by dimensional analysis alone and the detailed structure of the representation acting on it must be considered (similarly as in the case of algebras ${\bf A_{6,14}}$, ${\bf A_{6,18}}$ above).
\medskip

Let us now turn our attention to non--nilpotent radicals. Let us assume that $\r$ is a solvable Lie algebra with the nilradical $\n$. The existence of a Levi extension $\g$ of the non--nilpotent radical $\r$ with a Levi factor $\p$ imposes restrictions that go beyond the ones originating from the existence of $\g'=\n \dotplus \p$.
(On the other hand, we have already observed that $\g'$ may be decomposable and consequently not included in the lists in \cite{Tur1,Tur4} even if $\g$ is indecomposable.)

Let us assume that $\r=\n\dotplus\q$, $\ad(\p)|_{\q}=0$ as is always possible to achieve by the theorem of Turkowski. Then we have
$$ \ad(p)[x,y]=[\ad(p)x,y]+[x,\ad(p)y]=0$$
for any $x,y\in\q$ and $p\in\p$, i.e. the subspace $[\q,\q]\subset \n$ must be a representation space of the trivial representation (if $[\q,\q]$ is nonvanishing). 
Furthermore, for any $z\in\n,\, y\in\q,\, p\in\p$ we have
$$ \ad(p)[y,z]=[y,\ad(p)z]$$
i.e. similarly as in the proof of Theorem \ref{strLevnilp} we see that $\ad(y),\,y\in\q$ maps any representation subspace $V\subset \n$ of an irreducible representation of $\p$ either to a representation space of an equivalent representation (including $V$ itself) or to zero. 
 
Another restriction comes from the fact that $[\p,\r]\subset\n$ which for the corresponding derivations acting on $\n$ means that
$$ [\ad(p)|_{\n},\ad(x)|_{\n}]\in \inn(\n), \; \forall p\in\p,x\in\r.$$
This in turn implies that the $\m_1\m_1$--blocks of  $\ad(p)|_{\n}$ and $\ad(x)|_{\n}$ commute.

We collect some of these results into the following theorem
\begin{theorem}\label{sld}
Let $\g$ be a Levi decomposable Lie algebra which cannot be decomposed into a direct sum of ideals, $\p$ its Levi factor, $\r$ its radical, $\n$ its nilradical. Let $\n=\sum_{k=1}^K \m_k$ be the decomposition (\ref{nintoms}) of the nilradical. Then for any $p\in\p$ and $x\in\r$ the submatrices $(\ad(p))_{\m_1\m_1}$ and $(\ad(x))_{\m_1\m_1}$ of $\ad(p)|_{\n}$ and $\ad(x)|_{\n}$, respectively, commute
$$ [(\ad(p))_{\m_1\m_1},(\ad(x))_{\m_1\m_1}] =0.$$
In particular, if the restriction of $\ad(\p)$ to $\m_1$ is irreducible and $\g$ is an algebra over $\C$ then $\dim \r-\dim \n\leq 1$. When equality holds then the $\m_1\m_1$--block of the derivation $\ad(f_1)$ ($f_1\in \r \setminus \n$) is a nonvanishing multiple of the unit operator.
\end{theorem}
The proof of the statements in the particular case when $\ad(\p)|_{\m_1}$ is irreducible is a direct consequence of Schur's lemma. $\square$
\medskip

Theorem \ref{sld} can be used in explaining the particular values of parameters of solvable radicals $\r$ allowing Levi extension. E.g. in the algebra $g_{6,54}$ in \cite{Mub3} there are two parameters whereas its Levi extension $L_{9,49}^p$ in \cite{Tur4} has only one. The reason is that in order to have $(\ad(f_1))_{\m_1\m_1}$ commuting with  $(\ad(\p))_{\m_1\m_1}$ one of the parameters must be equal to 1. For the same reason the four parameters in the algebra ${\cal N}^{\alpha\beta\gamma\delta}_{6,1}$ of \cite{Tur2} are reduced to the values $\gamma=\alpha=p,\beta=\delta=q$ in the Levi extension $L_{9,28}^{p,q}$ of \cite{Tur4}. And similarly for other parametric families in Turkowski's classification \cite{Tur4}.

\section{Conclusions}
We have established an improved upper bound on dimension of any solvable extension of a given nilpotent Lie algebra. The new estimate (\ref{mybound})
$$  f \leq \dim \n - \dim \n^2, \qquad {\rm where} \quad f=\dim \s - \dim \n $$
 is different from the one derived by Mubarakzyanov in \cite{Mub1} 
\begin{equation}
 f \leq \dim \n 
\end{equation}
and improved in  \cite{Mub4} to 
\begin{equation}\label{Mubar_bound}
 f \leq \dim \n - \dim C(\s).
\end{equation}
There are at least two advantages to the estimate (\ref{mybound}) over (\ref{Mubar_bound}):
\begin{itemize}
 \item the bound (\ref{mybound}) is in most cases more restrictive than (\ref{Mubar_bound}),
 \item and it doesn't depend on the knowledge of the structure of the whole solvable Lie algebra $\s$, contrary to the bound (\ref{Mubar_bound}).
\end{itemize}
The bound (\ref{mybound}) is saturated for many classes of nilpotent Lie algebras whose solvable extensions were previously investigated -- e.g. Abelian \cite{NW}, naturally graded filiform $\n_{n,1}$, ${\cal Q}_n$ \cite{SW,ACSV}, a decomposable central extension of $\n_{n,1}$ in \cite{WLD} and triangular in \cite{TW}.

On the other hand, it is obvious that even the improved bound (\ref{mybound}) cannot give a precise estimate of the maximal dimension of a solvable extension in all cases. In particular, we have always $\dim \n - \dim \n^2\geq 2$, i.e. characteristically nilpotent Lie algebras cannot be easily detected using Eq. (\ref{mybound}). Similarly, the bound (\ref{mybound}) is not saturated in the case of Heisenberg nilradicals $\h$ \cite{RW} where the maximal number of non--nilpotent elements is in fact equal to $\frac{\dim \h+1}{2} < \dim \h -1$. It remains an open problem to further improve the estimate (\ref{mybound}) if it is possible.
\medskip

An interesting observation arises from investigation of numerous nilradicals in \cite{RW,NW,TW,SW,SW1,SK,ACSV,WLD,CS}. In all these cases the maximal solvable extension of the given nilradical over the field of complex numbers turns out to be unique up to isomorphism.
Therefore we formulate it as a conjecture
\begin{conjecture}
Let $\n$ be a complex nilpotent Lie algebra, not characteristically nilpotent. Let 
$\s$,$\tilde{\s}$ be solvable Lie algebras with the nilradical $\n$, of maximal dimension in the sense that no such solvable algebra of larger dimension exists. Then $\s$ and $\tilde{\s}$ are isomorphic. 
\end{conjecture}
It would be of interest to establish whether this conjecture holds in general or requires some supplementary assumptions on the structure of $\n$. In particular, if it holds for filiform nilradicals, then the classification of their solvable extensions becomes almost complete -- the filiform algebras which are neither characteristically nilpotent nor naturally graded would have precisely one solvable extension.
\medskip

Next, we have investigated the structure of Levi decomposable algebras. We have formulated several general properties that the nilradical of any Levi decomposable algebra must necessarily satisfy and applied these to the lists of Levi decomposable algebras in the papers \cite{Tur1,Tur4} by P. Turkowski. It turns out that dimensional analysis of the three characteristic series and their centralizers is enough to determine whether a given 5--dimensional nilpotent algebra has a nontrivial Levi extension. In the case of 6--dimensional nilpotent algebras this is no longer a sufficient criterion and more involved considerations were necessary. Nevertheless, we were able to explain the absence of a Levi extension for all but two 6--dimensional indecomposable nilpotent algebras by abstract, mostly dimensional, considerations, without the explicit construction of derivations. This indicates that techniques developed here can be of significant help in this kind of analysis in higher dimensions too.

The results and methods used in this section can be applied to Levi extensions of arbitrary dimension. One particular immediate consequence of them is that no filiform algebra can be a nilradical of a Levi decomposable algebra. More generally, the same holds also for any nilpotent algebra $\n$ such that $\dim\n-\dim\n^2=2$ and exists $j\in\N$ such that $\dim\n^j-\dim\n^{j+1}=\dim\n^{j+1}-\dim\n^{j+2}=1$. It remains an open problem to find some structurally interesting series of nilradicals in arbitrary dimension allowing the classification of its nontrivial Levi extensions other than $\n$ Abelian or Heisenberg.

\ack This research was supported by the research plan MSM6840770039 and the project LC527 
of the Ministry of Education of the Czech Republic. Discussions with Dietrich Burde, Dalibor Kar\'asek and Pavel Winternitz on the subject are gratefully acknowledged.

\end{document}